\documentclass[epj,nopacs]{svjour}
\pdfoutput=1
\usepackage{graphicx,hyperref,wrapfig,cite,color,subfigure,amssymb,amsmath,epsfig,
inputenc}
\hypersetup{
     colorlinks=true,       
     linkcolor=red,          
     citecolor=green,        
     filecolor=magenta,      
     urlcolor=blue           
}
\newcommand{\newc}{\newcommand}
\def\u#1{\verb!#1!\endgroup}

\newc{\HW}{\textsf{HERWIG}}
\newc{\Hw}{\textsf{Herwig}}
\newc{\TAUOLA}{\textsf{TAUOLA}}
\newc{\ThePEG}{\textsf{ThePEG}}
\newc{\boost}{\textsf{BOOST}}
\newc{\HepMC}{\textsf{HepMC}}
\newc{\Rivet}{\textsf{Rivet}}
\newc{\lhapdf}{\textsf{LHAPDF}}
\newc{\HWPP}{\textsf{Herwig++}}
\newc{\evt}{\textsf{EvtGen}}
\newc{\fortran}{\textsf{FORTRAN}}
\newc{\decayer}{\textsf{Decayer}}
\newc{\matchbox}{\textsf{Matchbox}}

\newc{\HWPPClass}[1]{\href{https://herwig.hepforge.org/doxygen/classHerwig_1_1#1.html}{\textsf{#1}}}
\newc{\ThePEGClass}[1]{\href{https://thepeg.hepforge.org/doxygen/classThePEG_1_1#1.html}{\textsf{#1}}}
\newc{\HWPPParameter}[2]{\href{https://herwig.hepforge.org/doxygen/#1Interfaces.html\##2}{{\bf #2}}}
\newc{\ThePEGParameter}[2]{\href{https://thepeg.hepforge.org/doxygen/#1Interfaces.html\##2}{{\bf #2}}}
\newc{\HWPPParameterValue}[3]{\href{https://herwig.hepforge.org/doxygen/#1Interfaces.html\##2}{{\bf [#2=#3]}}}
\newc{\HWPPParameterValueB}[3]{\href{https://herwig.hepforge.org/doxygen/#1Interfaces.html\##2}{{\bf [#3]}}}
\newc{\ThePEGParameterValue}[3]{\href{https://thepeg.hepforge.org/doxygen/#1Interfaces.html\##2}{{\bf [#2=#3]}}}

\preprint{
CERN-PH-TH-2015-289\\ 
MAN/HEP/2015/15\\ 
IFJPAN-IV-2015-13\\
HERWIG-2015-01\\
KA-TP-18-2015\\
DCPT/15/142\\
MCnet-15-28\\
IPPP/15/71}

\title{Herwig 7.0 / Herwig++ 3.0 Release Note}

\author{%
Johannes Bellm\inst{1,2} \and
Stefan Gieseke\inst{1} \and
David Grellscheid\inst{2} \and
Simon Pl\"atzer\inst{2,3} \and
Michael Rauch \inst{1} \and
Christian Reuschle\inst{1,4} \and
Peter Richardson\inst{2,5} \and
Peter Schichtel \inst{2} \and
Michael H. Seymour\inst{3} \and
Andrzej Si\'odmok\inst{5,6} \and
Alexandra Wilcock \inst{2} \and
Nadine Fischer \inst{1} \and
Marco A. Harrendorf \inst{7} \and
Graeme Nail\inst{3} \and
Andreas Papaefstathiou\inst{5} \and
Daniel Rauch \inst{1}
}

\institute{
Institute for Theoretical Physics, Karlsruhe Institute of Technology\and
IPPP, Department of Physics, Durham University\and
Particle Physics Group, School of Physics and Astronomy, University of Manchester\and
HEP Theory Group, Department of Physics, Florida State University\and 
CERN, PH-TH, Geneva\and
The Henryk Niewodniczanski Institute of Nuclear Physics in Cracow, Polish Academy of Sciences\and
Institut f\"ur Experimentelle Kernphysik, Karlsruhe Institute of Technology}

\date{\today}

\abstract{A major new release of the Monte Carlo event generator Herwig++
  (version 3.0) is now available. This release marks the end of distinguishing
  Herwig++ and HERWIG development and therefore constitutes the first major
  release of version 7 of the Herwig event generator family. The new version
  features a number of significant improvements to the event simulation,
  including: built-in NLO hard process calculation for all Standard Model
  processes, with matching to both angular-ordered and dipole shower modules
  via both subtractive (MC@NLO-type) and multiplicative (Powheg-type)
  algorithms; QED radiation and spin correlations in the angular ordered
  shower; a consistent treatment of perturbative uncertainties within the hard
  process and parton showering. Several of the new features will be covered in
  detail in accompanying publications, and an update of the manual will follow
  in due course. \PACS{ {xx.yy.zz}{Xx Yy Zz} } }

\begin{document}\sloppy

\maketitle

\section{Introduction}

\Hw\ is a multi purpose particle physics event generator. It is based on the
experience gained with both the \HW\ \cite{Corcella:2000bw} and \HWPP\
\cite{Bahr:2008pv} event generators. The latest version of \HWPP, 3.0,
marks the point at which the physics capabilities of the \HW\ version 6
series are fully superseded, and thus the last point at which their
development is distinguished. \HWPP\ 3.0 will henceforth be known as
\Hw\ 7.0. It replaces any prior \HW\ or \HWPP\ versions.

\Hw\ provides highly improved and extended
physics capabilities compared to both its predecessors, in particular
the ability to perform simulations at next-to-leading order in QCD,
while keeping the key physics
motivations such as coherent parton showers (including both angular-ordered and
dipole evolution), the cluster hadronization model, an eikonal multiple
interaction model and highly flexible BSM capabilities.

The last major public version (2.7) of \HWPP\ is described in great detail
in \cite{Bahr:2008pv,Bahr:2008tx,Bahr:2008tf,Gieseke:2011na,
  Arnold:2012fq,Bellm:2013lba}. This release note summarizes the major changes
and improvements introduced since then, which constitute the
base for the \Hw\ 7 series. The physics questions addressed by the
capabilities of \Hw\ 7 will be covered in detail in accompanying publications,
as well as comparisons with the other well-known general-purpose event
generators, Pythia\cite{Sjostrand:2007gs,Sjostrand:2014zea} and Sherpa\cite{Gleisberg:2008ta}.
A detailed manual covering all technical aspects will be prepared in
due course.
Please refer to \cite{Bahr:2008pv} and the present paper if using \Hw\ 7.0.

\subsection{Availability}

The new program version, together with other useful files and information, can
be obtained from the web site
\texttt{\href{https://herwig.hepforge.org/}{https://herwig.hepforge.org/}}. In
order to
improve our response to user queries, all problems and requests for user
support should be reported via the bug tracker on our wiki. Requests for an
account to submit tickets and modify the wiki should be sent to
\texttt{herwig@projects.hepforge.org}.

\Hw\ is released under the GNU General Public License (GPL) version 2 and
the MCnet guidelines for the distribution and usage of event generator
software in an academic setting, which are distributed together with the
source, and can also be obtained from
\texttt{\href{http://www.montecarlonet.org/index.php?p=Publications/Guidelines}
  {http://www.montecarlonet.org/}}

\subsection{Prerequisites}

\Hw\ 7.0 is based on \ThePEG\ 2.0, which is available along with the
\Hw\ installation sources at
\texttt{\href{https://herwig.hepforge.org/downloads}{https://herwig.hepforge.org/downloads}}. Further\linebreak
requirements are \boost\ \cite{boost}, \textsf{gsl}\ \cite{GSL},
\textsf{fastjet} \cite{Cacciari:2011ma} and \lhapdf\ \cite{Buckley:2014ana},
while a number of other dependencies are necessary in order to fully exploit
the program's capabilities. Amongst these are \HepMC\ and/or
\Rivet\ \cite{Buckley:2010ar} to analyze simulated events, as well as some or
all of the external amplitude libraries discussed in
section~\ref{sections:externals}.

In order to simplify the installation process, we provide a
\textsf{bootstrap} script to facilitate a consistent build and
installation of \Hw\ in a convenient way. The script requires a
\textsf{python} installation, and is available from
\texttt{\href{https://herwig.hepforge.org/herwig-bootstrap}
{https://herwig.hepforge.org/herwig-bootstrap}}.

\subsection{Documentation}

A significant new feature is the online documentation, which has been
completely rewritten and greatly extended to reflect the major changes
introduced with this version and
replaces the wiki pages. It can be found at
\texttt{\href{https://herwig.hepforge.org/tutorials/}
  {https://herwig.hepforge.org/tutorials/}}. An update of the more detailed
physics and manual will be made available in a similar format in due course.
Code snippets are provided for a wide variety of control functions for
easy inclusion into input files.
Detailed documentation of the source code and input file interfaces
generated with \textsf{doxygen} is available at
\texttt{\href{https://herwig.hepforge.org/doxygen/}
  {https://herwig.hepforge.org/doxygen/}}.

\section{NLO Event Simulation}

A key ingredient in the design and development of \Hw\ 7.0 was to provide
event simulation at next-to-leading order (NLO) accuracy in the strong
coupling {\it by default} for as many Standard Model processes as possible in
an automated way. The program, with the help of external libraries used for
amplitude calculation, is now able fully automatically to assemble NLO QCD
corrections to virtually all Standard Model processes, including matching to
both of its parton-shower algorithms \cite{Gieseke:2003rz,Platzer:2009jq}, via
methods inspired by either the MC@NLO\cite{Frixione:2002ik} or
Powheg\cite{Nason:2004rx} type algorithms, which we refer to as subtractive
and multiplicative matching, respectively.

Based on extensions of the previously developed \matchbox\
module\cite{Platzer:2011bc}, NLO event simulation is now possible
without the requirement of separately running external codes and/or
dealing with intermediate event sample files. Slight changes have been
made to improve \Hw's steering at the level of input files, and
significant improvements are provided to integration and unweighting,
including parallelization to meet the requirements of more complex processes.

\subsection{The Matchbox Module}

The design of the \matchbox\ modules closely resembles the structure
of the NLO QCD cross section calculated within a subtraction paradigm,
including the matching subtractions required to consistently combine
such calculations with parton showering downstream. Subtraction terms
are available in a flexible way, though only Catani-Seymour dipoles
\cite{Catani:1996vz,Catani:2002hc} are provided so far,
including both massless and massive QCD as well as the subtraction terms
required for supersymmetric QCD corrections.

Parton-shower matching subtractions are provided on an equally
flexible footing, including those required for the angular-ordered
shower \cite{Gieseke:2003rz}, the dipole shower \cite{Platzer:2009jq},
as well as matrix-element corrected showers forming the basis of
Powheg-type matching. For the latter, we provide
additional functionality to sample the matrix-element correction
Sudakov using the adaptive method outlined in \cite{Platzer:2011dr}. In
order to simplify the calculation of matching subtractions for the
angular-ordered shower, the kinematics reconstruction used to work out
the final shower kinematics has been changed to avoid additional
Jacobian factors when compared to the dipole parameterization in the case
of a single (or in general, the hardest) emission.

\subsection{External Amplitude Providers}
\label{sections:externals}

In order to set up the full calculation of a cross section, \matchbox\ requires
plugins to provide the respective tree and one-loop amplitudes. These plugins
can be interfaced either at the level of matrix elements squared (or tree-loop
interferences, respectively), or at the level of helicity, colour-ordered
subamplitudes with both trace- and colour flow bases provided within the
\matchbox\ core through adapted versions of the \textsf{ColorFull}
\cite{Sjodahl:2014opa} and \textsf{CVolver} \cite{Platzer:2013fha}
libraries\footnote{Other choices of colour bases are straightforward to
  implement through a very transparent interface.}.  While we provide built-in
amplitudes for a limited number of processes, the bulk of Standard Model
processes can be simulated using external amplitude plugins.

Based on extensions of the BLHA standard \cite{Binoth:2010xt,Alioli:2013nda}, \Hw\
currently supports interfaces to \textsf{GoSam} \cite{Cullen:2014yla},
\textsf{MadGraph} \cite{Alwall:2011uj}, \textsf{NJet} \cite{Badger:2012pg},
\textsf{OpenLoops}\cite{Cascioli:2011va} and \textsf{VBFNLO}
\cite{Arnold:2008rz,Baglio:2014uba}. Amplitudes for a limited number of LHC relevant
processes are directly provided along with the release, and amplitudes for
electroweak Higgs plus jets production are available from the
\matchbox\ plugin \textsf{HJets++}, which is available in the \texttt{Contrib}
section of the \Hw\ 7.0 release.

\subsection{Electroweak corrections to VV production}

Electroweak corrections to the production of heavy vector boson pairs have
been computed \cite{Bierweiler:2013dja,Bierweiler:2012kw} and implemented into
the program, as outlined in \cite{Gieseke:2014gka}.  The corrections are
applied as an event reweighting factor $K(\hat s, \hat t)$. In
order to apply this correction in a meaningful way one has to ensure that
additional QCD corrections are not too large and apply reasonable cuts to the
final state, as detailed in \cite{Gieseke:2014gka}.  The reweighting is
straightforward when applied together with Powheg-matched QCD corrections.  If
subtractive QCD matching is going to be applied one should rather apply the
matching on information extracted directly from the leptons in the final
state, this is detailed as an alternative method in \cite{Gieseke:2014gka}.  In
order to apply the method, one has to download grid files for the actual $K$
factors from a {\href{https://www.hepforge.org/archive/herwig/ewgrids/}
  {public archive}} at \texttt{hepforge}.

\section{Improvements to the Angular-Ordered Parton Shower}

  This release includes a number of improvements, which finally bring the
  default angular-ordered parton shower to the same level of accuracy as that in
  \HW\ 6.

\subsection{QED Showering}

The emission of QED radiation was not included in \HWPP. In \Hw\ 7.0 it is included in
the following way:
\begin{itemize}
\item A maximum scale is selected for QED radiation in the same way as for QCD radiation,
although selecting from the other charged particles in the process rather than the
colour partner in order to determine the scale. This scale need not be the same as the maximum scale for QCD
radiation.
\item Trial QCD and QED emissions are generated and the one with the
      higher scale selected, as required by the competition algorithm. This branching
      is generated as before and then any subsequent emissions of the same type
      are required to be angular 
      ordered\footnote{With the proviso discussed below for $g\to q\bar{q}$ branchings
      or in the the QED case $\gamma\to f\bar{f}$ splittings.} while those
      of a different type are only required to be ordered. For example
      if we generate a $q\to q g $ emission at an evolution scale $\tilde{q}_1$ and the
      quark has light-cone momentum fraction $z_1$ then any subsequent $q\to q g$ emissions
      must occur at a scale $\tilde{q}_2<z\tilde{q}_1$, as required by angular ordering. However any 
      QED $q\to q \gamma$ branchings need not be angular ordered and therefore
      can occur at an evolution scale $\tilde{q}_2<\tilde{q}_1$.
\end{itemize}

\subsection{Spin Correlations in the Shower}

There are correlations between the azimuthal angle of a branching and both the hard scattering process
and any previous branchings that occurred in the parton shower. There are two
types of correlation:
\begin{enumerate}
\item The soft correlation from the eikonal current, which correlates the direction of the
emitted gluon and the colour partner.
\item Spin correlations in the collinear limit between the azimuthal angle of the branching
and the hard process and any previous emissions.
\end{enumerate}
Both of these effects are included in \Hw\ 7.0 using the algorithm of
\cite{Knowles:1988vs,Knowles:1988hu,Collins:1987cp}. Now that the full spin correlations
are incorporated in the parton shower there is no requirement that unstable decays are generated before
the parton shower in order to generate the spin correlations between the production and decay of
the particles as described in \cite{Richardson:2001df,Gigg:2007cr}. The decays of unstable
fundamental particles are now handled as part of the parton-shower stage of the event
generation including all the spin correlations, both between the production of particles
and the parton-shower emissions, the production and decay of particles, and the
decay of particles and any parton-shower emissions.
The spin correlations are switched on by default and can be switched off using
\begin{small}\begin{verbatim}
set /Herwig/Shower/Evolver:SpinCorrelations No
\end{verbatim}\end{small}
While the soft correlations can be switched off using
\begin{small}\begin{verbatim}
set /Herwig/Shower/Evolver:SoftCorrelations No
\end{verbatim}\end{small}
we do not recommend this as the soft correlations affect the cluster mass spectrum and therefore
this change requires a retuning of the parton-shower and hadronization parameters.

As the spin correlations are currently not implemented in the shower subtraction terms used
at next-to-leading order the spin correlations are switched off by default when using NLO matching. 
However as we use the same formalism internally as \textsf{MadGraph} for the calculation
of helicity amplitudes \cite{Murayama:1992gi} the interface 
to \textsf{MadGraph} can fill the spin-density matrices used in the
spin-correlation algorithm and therefore the correlations can be correctly generated at leading
order.

\subsection[$g\to q\bar{q}$]{\boldmath{$g\to q\bar{q}$}}

The branching $g\to q\bar{q}$ is only singular in the collinear limit for massless quarks 
and does not have a soft singularity. It therefore should not be angular-ordered in the
parton shower, although given the nature of the parton shower algorithm it must continue to
be ordered in the evolution variable. We therefore relax the
constraint on this branching so that if a gluon is produced at a scale $\tilde{q}_1$ 
with light-cone momentum fraction $z_1$ the maximum scale of a subsequent $g\to q\bar q$
branching is now $\tilde{q}_1$, the maximum allowed by ordering of the evolution 
variable, rather than $z_1\tilde{q}_1$ as required by angular ordering. The
maximum evolution scale for other branchings remains unchanged.

Similarly the arguments presented in \cite{Bahr:2008pv, Buckley:2011ms} that
the scale used in the strong coupling for a branching should be the relative
transverse momentum, $p_\perp$, do not apply and therefore we have changed
this scale to be the invariant mass of the $q\bar{q}$ pair for this branching
only.\footnote{A similar change has been introduced in the dipole shower.}

\section{Perturbative and Shower Uncertainties}

Perturbative uncertainties in all of the hard processes provided by the
\matchbox\ module can be assessed by variation of the renormalization and
factorization scales, respectively. When fixed-order predictions at leading or
next-to-leading order are combined with subsequent parton showering, variations
of the renormalization and factorization scales in the parton shower ({\it
  i.e.}\ variations of the scale arguments of $\alpha_s$ and the parton
distribution functions) should be performed in a {\it correlated} way along
with variations in the hard process. While independent variations are
technically possible to assess patterns of scale compensation, the default
uncertainty settings will perform a consistent variation.

In addition to estimating unknown higher-order corrections by variation of the
renormalization and factorization scale, genuine parton-shower uncertainties
due to missing higher logarithmic orders and phase-space constraints can
also be estimated by varying the hard scale in the parton shower. There is no unique
definition of such a scale, and the relevant quantity is a specific detail of the
parton-shower algorithm and varies considerable between different approaches.
We provide variations of the relevant scale
in both the angular-ordered and dipole shower algorithms, which can be
used to assess these uncertainties, which are expected to be reduced by
use of NLO matched simulations. Specifically for this purpose, easily usable
settings of strict leading order simulation to be compared to improved NLO
simulation are provided within the new steering formalism summarized in
section~\ref{sections:steering}.

\section{Tuning}
\label{sections:tuning}

The improvements to both shower modules, as well as the inclusion of
next-to-leading order cross sections, have required a new tune to $e^+e^-$
data; this tune has been carried out using standard methods based on the
\textsf{Professor} framework \cite{Buckley:2009bj} using a representative set
of $e^+e^-$ data as previously described in \cite{Bahr:2008pv}. Similar
parameters and an overall reasonable description of the data have been
obtained for both the angular ordered and dipole shower. The results of these
tuning efforts are the default for the \Hw\ 7.0 release.

\subsection{Tuning of the Multi-parton interaction model}
It was shown in Ref.~\cite{Seymour:2013qka} that a good description of both underlying event 
and double parton scattering data \cite{Bahr:2013gkj} \emph{can} be obtained 
if one includes the latter in the data being fit to with a sufficiently high weight. 
We followed the procedure described in Ref.~\cite{Seymour:2013qka} using
the MMHT2014 LO 
parton distribution function~\cite{Harland-Lang:2014zoa}\footnote{In the near future we 
also plan to provide tunes using CT14~\cite{Dulat:2015mca} and NNPDF3.0~\cite{Ball:2014uwa}
parton distribution functions.} and obtained a tune consistent with double parton scattering 
data ($\sigma_{\!\textit{eff}}\approx15\,\mathrm{mb}$) that also gives a good
description of the underlying event data from the Tevatron's lowest
analysed energy
point\cite{Aaltonen:2015aoa}, $\sqrt{s}=300\,\mathrm{GeV}$ to the LHC's
highest\cite{Aad:2010fh}, $\sqrt{s}=7\,\mathrm{TeV}$.

\Hw\ 7.0 is released together with the tune \textsc{H7-UE-MMHT}, which
it uses by default.
More information and other related tunes can be obtained from
\href{https://herwig.hepforge.org/tutorials/mpi/tunes.html}{the \Hw\ tunes page}.

\section{Steering, Integration and Run Modes}
\label{sections:steering}

Owing to the complexity of the processes that can be simulated with
\Hw, this version introduces some new run modes as well as highly
simplified input files to ease steering the event generator. Two alternative
integrator modules are provided in addition to the old default, \textsf{ACDC} of
\ThePEG, providing superior performance especially for more complex
processes. One of the algorithms is based on the standard sampling algorithm
contained in the \textsf{ExSample} library \cite{Platzer:2011dr}, while
the other is based on the \textsf{MONACO} algorithm, a \mbox{\textsf{VEGAS}}
\cite{Lepage:1980dq} variant, used by \textsf{VBFNLO}
\cite{Arnold:2008rz,Baglio:2014uba}.

Since both of these algorithms require an integration grid to be set up prior to
generating events, two levels of run mode have been introduced in
addition to the old \texttt{read} and \texttt{run} steps, to meet the
requirements of
more complex processes. The new
\texttt{integrate} step performs the grid adaptation; it is possible to
parallelize this step in a way that does {\it not\/} require inter-process
communication and the individual tasks in this parallelization can easily be
submitted to standard batch or grid queues. The \texttt{integrate} step is to be
preceded by a \texttt{build} step\footnote{The old \texttt{read} step is still
  available, representing the subsequent execution of both the \texttt{build}
  and \texttt{integrate} steps in one step.}, which will assemble the full
fixed-order or matched cross section, including subtraction terms and the
possibility of external amplitude libraries generating dedicated code for the
process of interest. As this step may also require considerable
computational resources, the \texttt{integrate} and \texttt{run} steps both support
reading in additional input files, so-called setup files, to modify run
parameters independently of the process of building an event generator object.
Detailed examples of these various new work-flows are given in the
\href{https://herwig.hepforge.org/tutorials/}
{new documentation}.

Event generation itself can be parallelized either through submitting
runs with explicitly set random seeds or through the newly introduced
feature of forking several
event generation jobs on multicore nodes.

\section{Herwig Contrib Projects}

A number of related codes have been developed along with the main \Hw\ 7.0
development; while these libraries are not supported at the same level as the
core \Hw\ release, they are provided along with it. Amongst
other tools, the new program version provides the following
plugins:

\subsection{Electroweak Higgs plus Jets Production}

A dedicated \matchbox\ plugin providing amplitudes for the calculation of
electroweak Higgs plus jets production at NLO QCD is available along with the
release. This library has been used in the calculation reported in
\cite{Campanario:2013fsa}. It provides a full calculation of $pp\to h + n$
jets at ${\cal O}(\alpha^3\alpha_s^{n-2})$ for $n=2,3,4$ at leading, and
$n=2,3$ at next-to-leading order QCD. All relevant topologies of either VBF or
Higgs-Strahlung type are taken into account along with all interferences. The
technical details of the library will be described elsewhere; its use is the
same as for all other \matchbox-based calculations and a corresponding input
file snippet to enable this class of processes is provided.

\subsection{FxFx merging support}

\Hw\ 7.0 contains interface support for FxFx merging
\cite{Frederix:2012ps}, a method for merging multi-jet
NLO samples with a parton shower. The
interface allows usage of samples generated from \texttt{MadGraph
  5/aMC@NLO}~\cite{Alwall:2014hca}. The module has been tested for $\mathrm{W+jets}$
and $\mathrm{Z+jets}$ events, and compared against LHC data at 7 and 8 TeV~\cite{Frederix:2015eii}. Other
processes will be supported in future releases.

\subsection{Higgs boson Pair Production}

The \textsf{HiggsPair} and \textsf{HiggsPairOL} packages offer production of
Higgs boson pairs via gluon fusion. The former uses code from
\textsf{HPAIR}~\cite{Dawson:1998py, Plehn:2005nk} whereas the latter
uses the \textsf{OpenLoops} one-loop generator for the matrix
elements~\cite{Cascioli:2011va}.

\textsf{HiggsPair} describes leading-order Higgs boson pair
production, either in the Standard Model or in its $D=6$ effective field
theory extension. The original implementation was described
in~\cite{Bellm:2013lba} and its $D=6$ EFT extension was
examined in detail in~\cite{Goertz:2014qta}.

\textsf{HiggsPairOL} describes SM Higgs boson pair production, with
the optional use of Higgs-Higgs+one jet matrix elements merged to the parton
shower via the MLM method. See~\cite{Maierhofer:2013sha} for a detailed
description.

\section{Sample Results}

\begin{figure}[htbp]
  \centerline{\resizebox{\columnwidth}{!}{\includegraphics{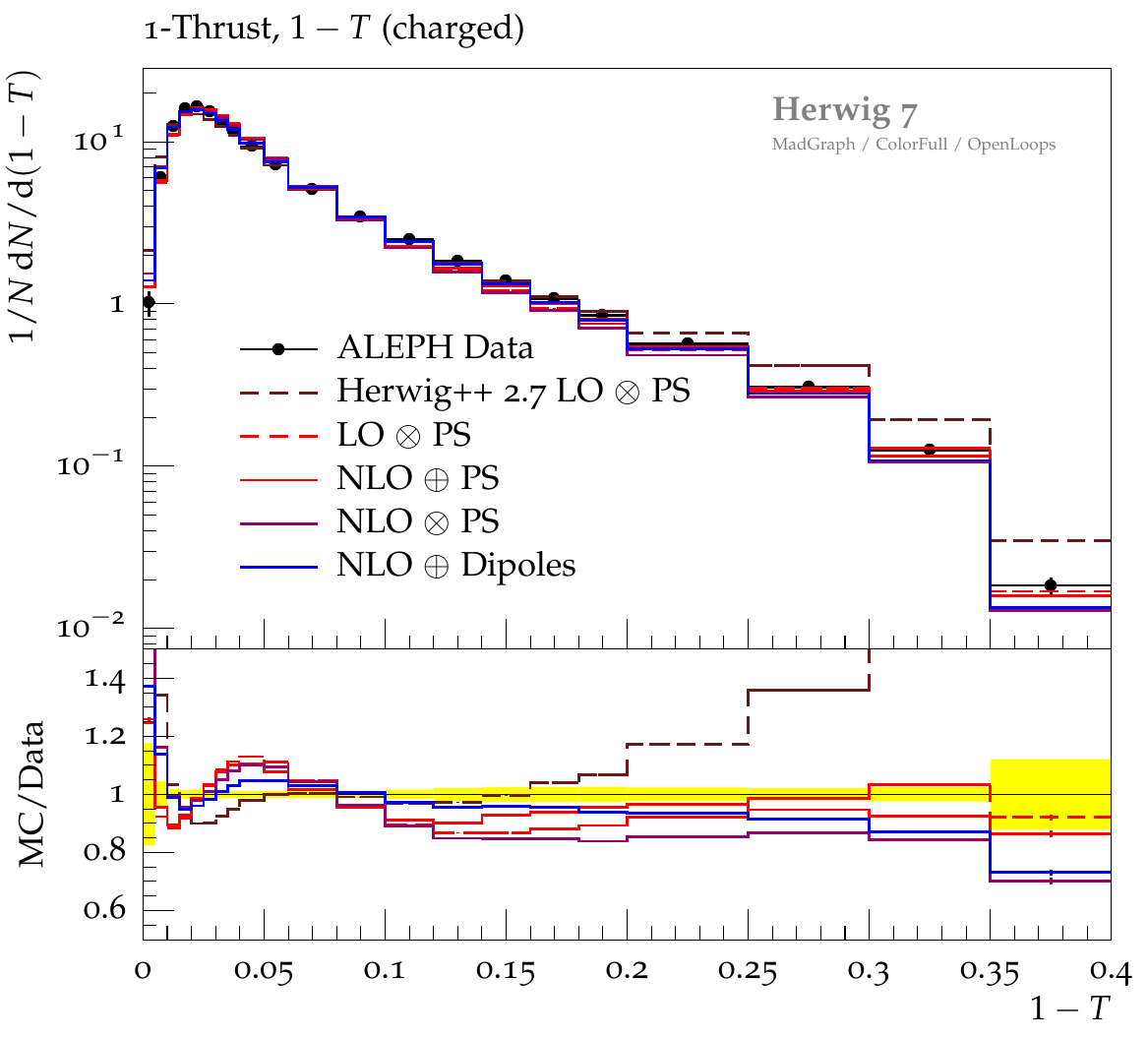}}}
  \caption{The thrust distribution in $\mathrm{e^+e^-}$ annihilation at
    $\sqrt{s}=M_z$, in comparison with ALEPH data\cite{Barate:1996fi}.}
  \label{results1}
\end{figure}

\begin{figure}[htbp]
  \centerline{\resizebox{\columnwidth}{!}{\includegraphics{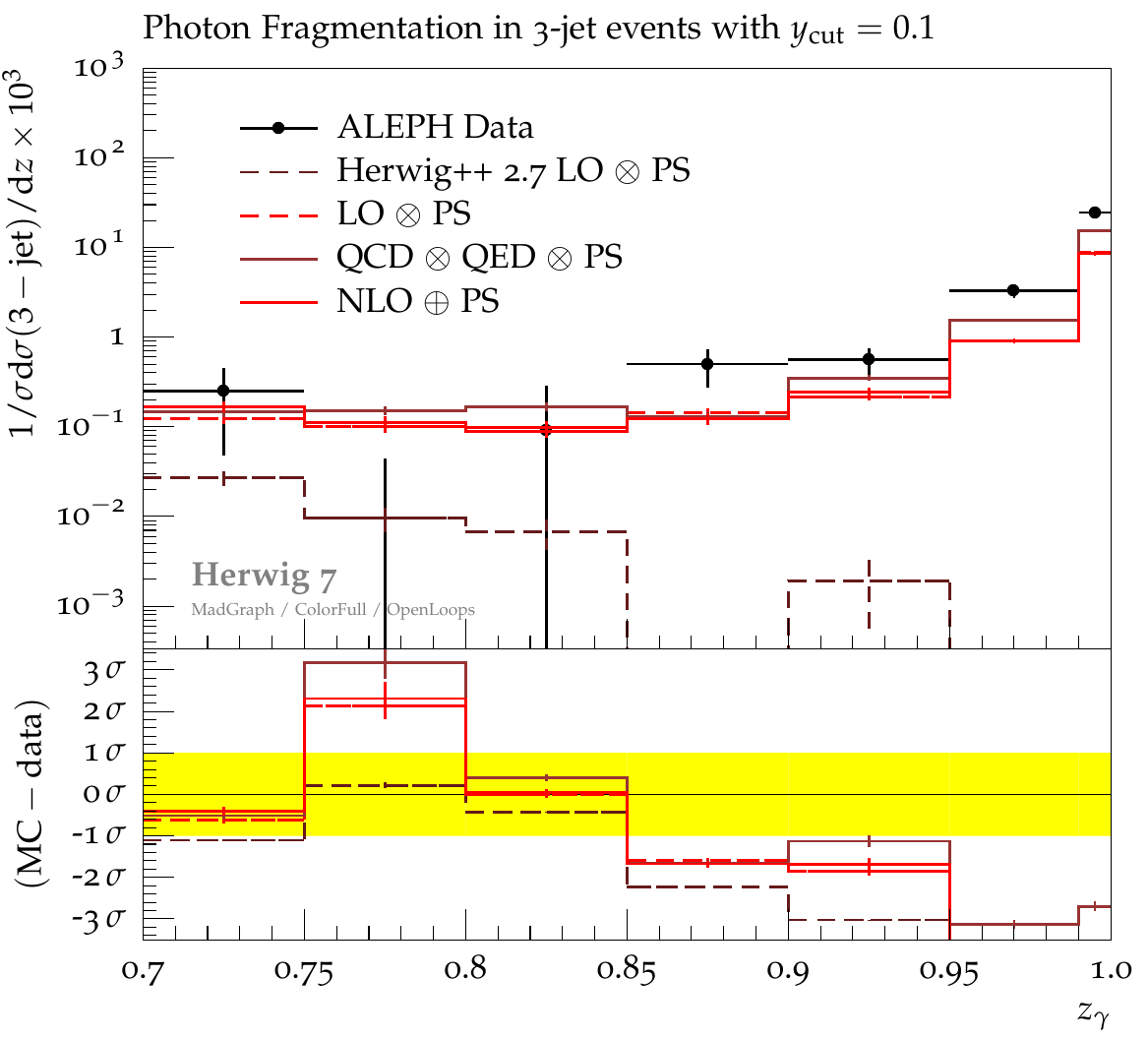}}}
  \caption{The distribution of photon-jet energy fraction in three-jet
    $\mathrm{e^+e^-}$ events at $\sqrt{s}=M_z$ defined with a cutoff in
    the $k_\perp$ algorithm of $y=0.1$ in comparison with ALEPH
    data\cite{Buskulic:1995au}.}
  \label{results2}
\end{figure}

\begin{figure}[htbp]
  \centerline{\resizebox{\columnwidth}{!}{\includegraphics{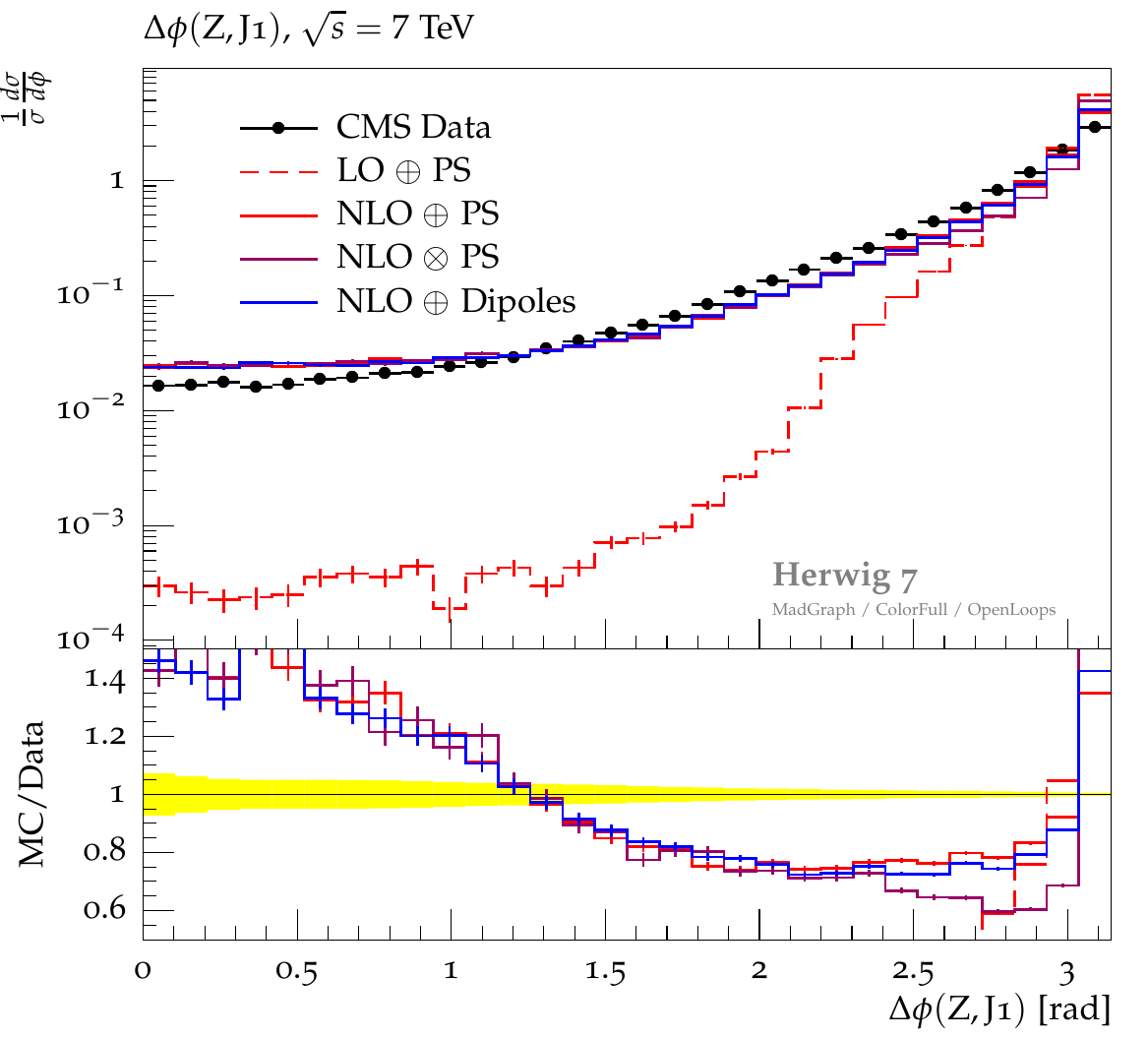}}}
  \caption{The distribution of separation in azimuthal angle between
    the $\mathrm{Z}$ boson and the hardest jet in $\mathrm{Z+jets}$
    events in $\mathrm{pp}$ collisions at $\sqrt{s}=7\,\mathrm{TeV}$ in
    comparison with CMS data\cite{Chatrchyan:2013tna}.}
  \label{results3}
\end{figure}

\begin{figure}[htbp]
  \centerline{\resizebox{\columnwidth}{!}{\includegraphics{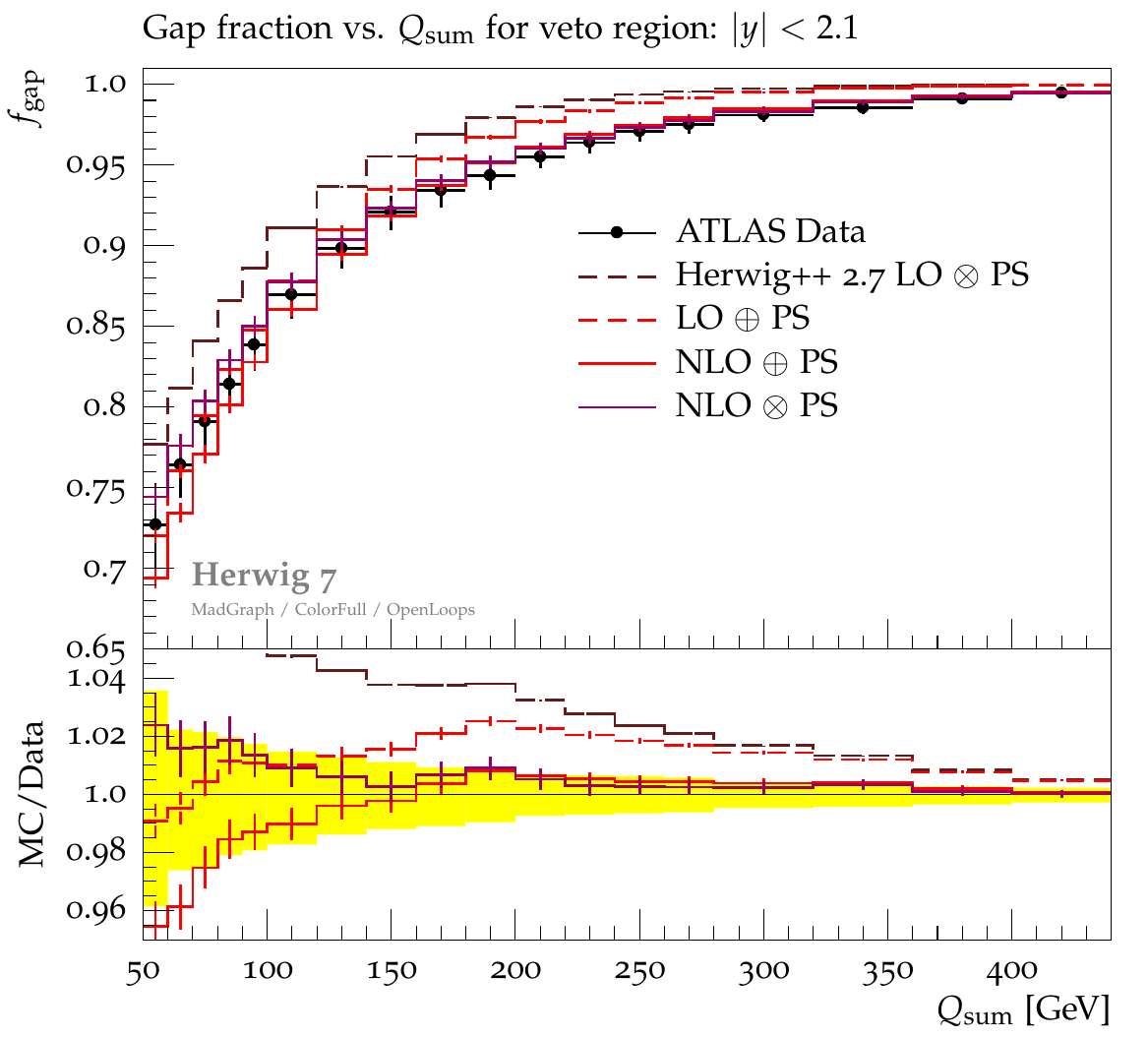}}}
  \caption{The fraction of events that have less than $Q_{\mathrm{sum}}$
    transverse energy in the rapidity region $|y|<2.1$ in top
    quark-antiquark events in $\mathrm{pp}$ collisions at
    $\sqrt{s}=7\,\mathrm{TeV}$ in comparison with ATLAS
    data\cite{ATLAS:2012al}.}
  \label{results4}
\end{figure}

With so many new features, it is impossible to show the full spectrum of
results that have been improved, but in
figures~\ref{results1}--\ref{results4} we show a small sample.

The Monte Carlo results shown are from \HWPP\ version 2.7 using leading order
plus parton shower simulation and from \Hw~7.0 with the angular-ordered parton
shower (LO $\oplus$ PS), the angular-ordered parton shower supplemented by the
internally-implemented Powheg correction, which includes QCD and QED
corrections for the case of $e^+e^-\to q\bar{q}$ (QCD $\otimes$ QED $\otimes$
PS), by the automatically-calculated by Matchbox subtractive (MC@NLO-type)
matching (NLO $\oplus$ PS) and multiplicative (Powheg-type, NLO $\otimes$ PS)
corrections and, finally, the dipole shower supplemented by a subtractive
matching to NLO cross sections (NLO $\oplus$ Dipoles).

In Fig.~\ref{results1}, we show the most well-studied event shape from
the LEP era, the thrust distribution, in comparison with data from the
ALEPH collaboration\cite{Barate:1996fi}. A long-standing problem of
\HWPP\ producing too many very hard events, whether or not NLO matching
was used, is seen to have been solved by the improvements to the
angular-ordered shower algorithm. All of the variants of NLO matching
then give a similar description of the data, with the dipole shower
giving a somewhat better overall description.

In Fig.~\ref{results2}, the effect of the inclusion of photon emission in the
angular-ordered parton shower is shown. Events at $z_\gamma=1$ are isolated
photons (``jets'' for which all of the jet energy is carried by a single
photon), while events at lower $z_\gamma$ come from hard collinear photon
emission from the final state quark jets. We see clearly that the results from
\HWPP\ have no component at large $z_\gamma$ at all, while all of the \Hw\ 7.0
variants are much closer to the data with that including matching to NLO QED as well as QCD giving the best agreement. QED radiation within the dipole shower
is subject to ongoing development and will be available in a future
release.

In Fig.~\ref{results3} we turn to results for $\mathrm{Z+jets}$ events
at the LHC. We show the distribution of separation in azimuthal angle
between the $\mathrm{Z}$ boson and the hardest jet. The region
$\Delta\phi\sim\pi$ corresponds to leading order kinematics, in which
the $\mathrm{Z}$ boson gains its transverse momentum by recoiling
against a single hard parton, whereas the broad spectrum of events down
to $\Delta\phi=0$ corresponds to events in which the $\mathrm{Z}$ boson
recoils against two or more jets. The need for NLO corrections is
clearly seen. An important cross-check of the two different automated
NLO matching schemes and the two different shower algorithms both using
subtractive matching can also be seen.

Finally, in Fig.~\ref{results4} we show the jet activity in
$\mathrm{t\bar{t}}$ events at the LHC, as revealed by the gap fraction,
i.e.\ the fraction of events for which the sum of the transverse momenta
of all additional jets in the prescribed rapidity region is less than
$Q_{\mathrm{sum}}$. \HWPP\ 2.7 is seen to have far too little jet
activity (too many gap events). While \Hw\ 7.0 with the shower alone is
somewhat closer to the data at small $Q_{\mathrm{sum}}$, a clear deficit
is seen for hard jet events at high $Q_{\mathrm{sum}}$, while both the
NLO matching schemes describe the data well.

This is of course just a very small selection of the large number of distributions
that have been checked against data in the final preparations of
\Hw\ version 7.0, and more will be shown for specific processes in a series
of forthcoming papers.

\section{Summary and Outlook}

We have presented version 7.0 of the \Hw\ event generator, based on previous
\HWPP\ development and the experience gained with the \HW\ event
generator. The new program features significant improvements as compared to
both the \HWPP\ 2.x series and the \HW\ 6 event generator, amongst them
a powerful framework for NLO calculations and a number of improvements to both
shower modules. Several accompanying publications containing detailed
coverage of both physics and technical aspects will follow in due course, as
well as an updated large and detailed manual to replace
\cite{Bahr:2008pv}. A completely new documentation system is already in place
for \Hw\ 7 to allow the user to exploit the full capability of the new
program.  The methods and code developed within this release will also form
the basis for ongoing and future development such as multijet merging at both
leading and next-to-leading order, and electroweak corrections.

\section*{Acknowledgements} 

We are grateful to the authors of the predecessor \HW\ and \HWPP\ programs
for their many inputs into the development of \Hw\ and to the many users
whose feedback has led to its improvement. We are especially indebted to
Leif L\"onnblad for his authorship of \ThePEG, on which \Hw\ is built,
and his close collaboration, and also to the authors of \Rivet\ and
\textsf{Professor}.
We would like to thank Terrance Figy for numerous discussions, testing and
feedback. We are also grateful to Malin Sj\"odahl for providing the
\textsf{ColorFull} library for distribution along with \Hw.

This work was supported in part by the European Union as part of the FP7
Marie Curie Initial Training Network MCnetITN (PITN-GA-2012-315877).

It was also supported in part by the Lancaster-Manchester-Sheffield
Consortium for Fundamental Physics under STFC grant ST/L000520/1 and the 
Institute for Particle Physics Phenomenology under STFC grant ST/G000905/1.

AP and SP acknowledge support by FP7 Marie Curie Intra European Fellowships
(PIEF-GA-2013-622071
and PIEF-GA-2013-628739 respectively). CR acknowledges support by the
German Federal Ministry of Education and Research (BMBF).

SP is grateful to DESY Hamburg and KIT for kind hospitality while part of this
work was completed, as well as for providing computing resources. We are
also grateful to the Cloud Computing for Science and Economy project (CC1) at
IFJ PAN (POIG 02.03.03-00-033/09-04) in Cracow and the U.K. GridPP project
whose resources were used to
carry out some of the numerical calculations for this project.  Thanks also to
Mariusz Witek and Mi\l osz Zdyba\l\ for their help with CC1 and Oliver Smith for his help
with grid computing.

\bibliography{Herwig}
\end{document}